\documentclass[conference,10pt]{IEEEtran}

\usepackage{amsmath,amsxtra,amssymb,latexsym,amscd,amsfonts,mathrsfs}
\usepackage{verbatim} 
\usepackage{graphicx, epstopdf}
\usepackage{ float}
\usepackage[noadjust]{cite}
\usepackage{inputenc}
\usepackage[mathcal]{euscript}
\usepackage[noadjust]{cite}

\makeatletter
\newcommand*\titleheader[1]{\gdef\@titleheader{#1}}
\AtBeginDocument{%
  \let\st@red@title\@title
  \def\@title{%
    \bgroup\normalfont\small\centering\@titleheader\par\egroup
    \vskip0.2em\st@red@title}
}

\ifCLASSINFOpdf
\else
\fi

\title{Cooperation in NOMA Networks Under Limited User-to-User Communications: Solution and Analysis}

\begin{document}

\makeatother
\titleheader{This is the authors'version of the paper that has been accepted for publication in IEEE Wireless Communications and Networking Conference, 15-18 April 2018, Barcelona, Spain}
%




%

\author{\IEEEauthorblockN{ 
Duc-Dung Tran\IEEEauthorrefmark{1},
Ha-Vu Tran\IEEEauthorrefmark{2}, 
Dac-Binh Ha\IEEEauthorrefmark{1}, and
Georges Kaddoum\IEEEauthorrefmark{2}
}
\IEEEauthorblockA{\IEEEauthorrefmark{1}Faculty of Electrical and Electronics Engineering, Duy Tan University
Danang, Vietnam\\ Email: dung.td.1227@gmail.com, hadacbinh@duytan.edu.vn}
\IEEEauthorblockA{\IEEEauthorrefmark{2}ETS Engineering School, University of Quebec, Montreal, Canada \\
Email: ha-vu.tran.1@ens.etsmtl.ca, georges.kaddoum@etsmtl.ca}
}


\maketitle

\begin{abstract}
This paper proposes a new communication protocol for a cooperative non-orthogonal multiple access (NOMA) system. In this system, based on users' channel conditions, each two NOMA users are paired to reduce system complexity. In this concern, the user with a better channel condition decodes and then forwards messages received from the source to the user with a worse channel condition. In particular, the direct link between the paired users is assumed to be unavailable due to the weak transmission conditions. To overcome this issue, we propose a new cooperative NOMA protocol in which an amplify-and-forward (AF) relay is employed to help the user-to-user communications. To evaluate the proposed protocol, the exact closed-form expressions of outage probability (OP) at the two paired users are derived. Based on the analysis of the OP, we further examine the system throughput in a delay-sensitive transmission mode. Finally, our analytical results verified by Monte-Carlo simulation show that the proposed protocol is efficient in enhancing the performance of NOMA system when the user-to-user communications is limited.
\end{abstract}


\section{Introduction}
The fifth generation (5G) networks are expected to support multimedia applications to achieve a 1000-fold higher throughput, a 1000-fold higher mobile data per unit area, and a 10-fold longer lifetime of devices over the fourth generation (4G) networks \cite{AGupta2015,Ekram2015,Tran2017}. To reach these goals, in the quest for new technologies, non-orthogonal multiple access (NOMA) has emerged as one of the promising candidates \cite{Sai2013,Dai2015,Nak2015}. In fact, this method can be considered as a key solution to improve spectral efficiency. Specifically, its principle relies on exploring the power domain and users' channel conditions to serve multiple users at the same time/frequency/code \cite{Dai2015}. Furthermore, compared with conventional multiple access, NOMA can offer a better user fairness since the users with a weak transmission condition can be served at a timely manner \cite{Dai2015,Nak2015}.

On the other hand, cooperative communication is an outstanding solution to improve system performance and extend coverage areas \cite{Sur2009} for wireless networks. Particularly, a combination between cooperative transmission and NOMA has gained significant attention from the research community \cite{Kim2015,Lia2017,Men2015,Do2017,Din2015,Liu2016}. In this research line, the paper \cite{Kim2015} has investigated NOMA in cooperative networks in which the system consists of one base station (BS) and two users. In this concern, the BS communicates directly with the first user while the second one exchanges information with the BS through the help of a decode-and-forward (DF) relay. The authors have shown that the performance of outage probability and ergodic sum capacity is significantly improved by using the proposed NOMA. As an extension of \cite{Kim2015}, the work \cite{Lia2017} has considered a downlink cooperative NOMA system with the aid of an amplify-and-forward (AF) relay. This study has compared the overall outage probability of the cooperative NOMA with the conventional cooperative OMA to clarify the benefits of cooperative NOMA scheme. Moreover, the derivation of outage probability, diversity order and coding gain have been presented. Furthermore, in \cite{Men2015}, a NOMA-based downlink cooperative cellular system has been examined, in a scenario where the BS communicates with two paired mobile users via the help of a half-duplex amplify-and-forward (AF) relay. In addition, the work \cite{Do2017} has analyzed the outage performance of NOMA networks with cooperative relaying transmission, in which the energy-limited near users are powered by the source applying simultaneous wireless information and power transfer (SWIPT).
The authors have identified that NOMA can provide an improved spectral efficiency and user fairness in cooperative networks. 
In particular, the works in \cite{Din2015} and \cite{Liu2016} have exploited this advantage to improve the overall performance of cooperative communication. Specifically, they have proposed cooperative NOMA transmission schemes to improve the outage performance of the users with poor channel conditions by considering the users with good channel gains as relays to help the others.

In cooperative NOMA networks, the user with a better channel condition is responsible for decoding and then forwarding the messages to the user with a poorer condition \cite{Din2017}. In other words, it can be seen as a decode-and-forward (DF) relay.
On this basis, one can be observed that the user-to-user communication plays a critical role in the operation of the networks.
In fact, most of the previous works \cite{Kim2015,Lia2017,Men2015,Do2017,Din2015,Liu2016} consider their proposed systems under the assumption that the direct links between the users with good channel conditions. Nevertheless, in practice, these connections may be unavailable due to the weak transmission conditions or obstacles between the users. To our best's knowledge, there has been little work on such an issue.

Motivated by the above discussions, in this paper, we focus on designing a cooperative NOMA protocol to deal with limited user-to-user communications and enhance the reliability for the considered system. Thus, this scheme will be suitable to the networks with high reliability, such as Vehicle-to-anything (V2X) system \cite{Man2016}.  More specifically, in the proposed protocol, multiple users are divided into multiple pairs to perform cooperative NOMA network. Indeed, this manner helps to reduce an amount of system overhead, as well as used time slots, in comparison with combining all users to perform cooperative NOMA \cite{Din2017}. Considering the two paired users, conventionally, the user with a better channel gain works as a DF relay to enhance the quality of the received signal at the remaining user.
However, since the direct communication between the two paired users is unavailable due to a poor transmission condition or obstacles, the use of an AF relay is proposed to help the user with a better channel condition forward the signals to the user with a severe channel quality. 
The main contributions of our paper are presented as follows
\begin{itemize}
\item Proposing a cooperative NOMA protocol addressing the issue of limited user-to-user communications.
\item Deriving closed-form expressions of outage probability (OP) and system throughput for the considered system.
\item Exploring the impact of the distances between the communication nodes and users-paired selection, on the system performance.
\end{itemize}

Particularly, by comparing the performance of cooperative communication with that of  non-relaying communication, numerical results clarify the advantage of the proposed cooperative communication in NOMA networks.


\section{System model}

\begin{figure}[!t]
\centering
\includegraphics[scale = 0.3]{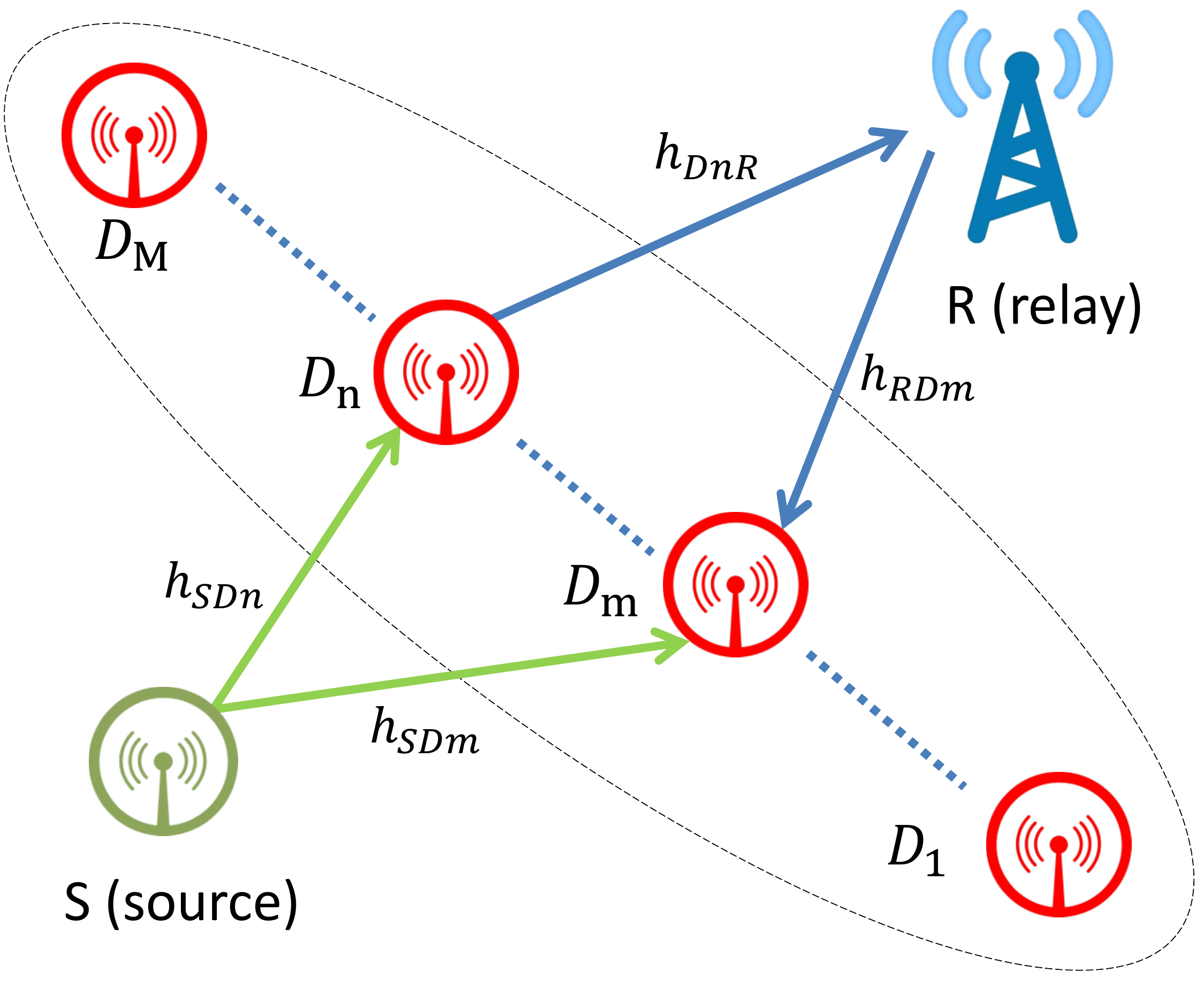}
\caption{ Model of cooperative NOMA system} 
\label{fig_model}
\end{figure}

As depicted in Fig. 1, we consider a downlink cooperative NOMA system. The considered network consists of one BS denoted by $S$, $M$ users denoted by $\{D_i\}$ $(1 \le i \le M)$, and one AF relay node denoted by $R$. 
Moreover, it is also assumed that each user has a single antenna and operates in a half-duplex mode. The channels are supposed to undergo frequency non-selective Rayleigh block fading.  In addition, the channel gains between BS $S$ and users $\{D_i\}$ are assumed to follow the order of $|{h_{SD_1}}{|^2} \le |{h_{SD_2}}{|^2} \le  \ldots  \le |{h_{SD_M}}{|^2}$. 

In the model, BS $S$ intends to convey information to users $\{D_i\}$. Prior to transmission, two users, e.g. $D_m$ and $D_n$ $(1 \le m < n \le M)$, are selected to perform NOMA. It is assumed that the direct link between $D_n$ and $D_m$ is unavailable due to the poor transmission conditions. On this basis, the proposed scenario can be described as follows
\begin{itemize}
\item In the first phase, source $S$ starts with transmitting the superimposed message to users $D_m$ and $D_n$ by applying NOMA.
\item In the second phase, given the two paired users, the user with the better channel condition, i.e. user $D_n$, first decodes the message of the remaining user, i.e user $D_m$, from the received signal, then performs successive interference cancellation (SIC) to remove this component from its observation and finally recovers its own information.
\item In the third phase, after decoding the message of user $D_m$, user $D_n$ forwards the result to user $D_m$ via the help of relay $R$. Thus, user $D_m$ receives two messages transmitted from both the source and relay. Specifically, selection combining (SC) scheme is employed at user $D_m$ to process these signals.
\end{itemize}


Given the proposed scenario, mathematical formulation can be provided as follows. In the first phase, $S$ transmits the message $x_s = \sqrt{a_m P_0}s_m + \sqrt{a_n P_0}s_n$ $(a_m > a_n)$ to two selected users $D_m$ and $D_n$ following NOMA. Specifically, $s_m$ and $s_n$ are the messages of users $D_m$ and $D_n$, respectively. Also, $a_m$ and $a_n$ are the power allocation coefficients satisfied the condition $a_m + a_n = 1$, and $P_0$ is the transmit power. Accordingly, the received signals at users $D_m$ and $D_n$ are respectively given by
\begin{equation}
y_{Dm} = \frac{{{h_{SDm}}}}{{\sqrt {d_{SDm}^\theta } }}\left( {\sqrt {{a_m}{P_0}} {s_m} + \sqrt {{a_n}{P_0}} {s_n}} \right) + {n_{Dm}},
\label{ydm}
\end{equation}
\begin{equation}
y_{Dn} = \frac{{{h_{SDn}}}}{{\sqrt {d_{SDn}^\theta } }}\left( {\sqrt {{a_m}{P_0}} {s_m} + \sqrt {{a_n}{P_0}} {s_n}} \right) + {n_{Dn}},
\end{equation}
where $d_{SDm}$ and $d_{SDn}$ are the distances from BS $S$ to users $D_m$ and $D_n$, respectively. Additionally, $\theta$ is the path loss exponent, and $n_{Dm}$ and $n_{Dn} \sim {\cal{CN}}(0,N_0)$ denote the additive white Gaussian noise (AWGN) at users $D_m$ and $D_n$, respectively. Thus, the instantaneous signal-to-interference-and-noise ratio (SINR) at user $D_m$ to detect $s_m$ is written as
\begin{equation}
{\gamma _{SDm}} = \frac{{{a_m}{{\left| {{h_{SDm}}} \right|}^2}}}{{{a_n}{{\left| {{h_{SDm}}} \right|}^2} + d_{SDm}^\theta /{\gamma _0}}},
\label{gsdm}
\end{equation}
where $\gamma_0 = \frac{P_0}{N_0}$ denotes the average transmit signal-to-noise ratio (SNR) at BS $S$.

In the second phase, user $D_n$ decodes the message of user $D_m$ (i.e. $s_m$), and then employs SIC to subtracts the signal $s_m$ from the received signal before decoding its own message (i.e. $s_n$). Right after, it forwards $s_m$ to relay $R$. It is assumed that the information processing times at user $D_n$ and relay $R$ are negligible and ignorable, respectively. The received signal at relay $R$ can be expressed as
\begin{equation}
y_R = \frac{{{h_{DnR}}}}{{\sqrt {d_{DnR}^\theta } }}\sqrt {{P_{Dn}}} {s_m} + {n_R},
\end{equation}
where, $h_{DnR}$ is the channel coefficient of $D_n - R$ link, $P_{Dn}$ is the transmit power at user $D_n$, $d_{DnR}$ is the $D_n - R$ distance, $n_R \sim {\cal{CN}}(0,N_0)$ is the AWGN at relay $R$. The instantaneous SINR at user $D_n$ to detect $s_m$ of user $D_m$ can be given by
\begin{equation}
{\gamma _{SD{n \to m}}} = \frac{{{a_m}{{\left| {{h_{SDn}}} \right|}^2}}}{{{a_n}{{\left| {{h_{SDn}}} \right|}^2} + d_{SDn}^\theta /{\gamma _0}}}.
\label{gsdnm}
\end{equation}
The instantaneous signal-to-noise ratio (SNR) at user $D_n$ to detect $s_n$ of user $D_n$ is written as
\begin{equation}
{\gamma _{SDn}} = \frac{{{\gamma _0}{a_n}{{\left| {{h_{SDn}}} \right|}^2}}}{{d_{SDn}^\theta }},
\label{gsdn}
\end{equation}

In the third phase, relay $R$ amplifies the received signal, and then re-transmits the result to user $D_m$. Hence, the received signal at user $D_m$ has the following form
\begin{equation}
\begin{split}
&y_{RDm} \\
&= \sqrt {\frac{{{P_R}{P_{Dn}}}}{{\left( {\frac{{{P_{Dn}}}}{{d_{DnR}^\theta }}{{\left| {{h_{DnR}}} \right|}^2} + {N_0}} \right)d_{RDm}^\theta d_{DnR}^\theta }}} {h_{RDm}}{h_{DnR}}{s_m} \\
& \quad + \sqrt {\frac{{{P_R}}}{{\left( {\frac{{{P_{Dn}}}}{{d_{DnR}^\theta }}{{\left| {{h_{DnR}}} \right|}^2} + {N_0}} \right)d_{RDm}^\theta }}} {h_{RDm}}{n_R} + {n_{RDm}},
\end{split}
\end{equation}
where $h_{RDm}$ is the channel coefficient of $R - D_m$ link, $P_{R}$ is the transmit power at relay $R$, $d_{RDm}$ is the $R - D_m$ distance, $n_{RDm} \sim {\cal{CN}}(0,N_0)$ is the AWGN at user $D_m$. Here, for simplicity but without loss of generality, we assume that $P_{Dn} = P_R = P_0$. The instantaneous SINR of user $D_m$ related to $R - D_m$ link is given by
\begin{equation}
\begin{split}
&{\gamma _{RDm}} \\
&= \frac{{\gamma _0^2{{\left| {{h_{RDm}}} \right|}^2}{{\left| {{h_{DnR}}} \right|}^2}}}{{{\gamma _0}d_{DnR}^\theta {{\left| {{h_{RDm}}} \right|}^2} + {\gamma _0}d_{RDm}^\theta {{\left| {{h_{DnR}}} \right|}^2} + d_{RDm}^\theta d_{DnR}^\theta }}.
\label{grdm}
\end{split}
\end{equation}

In order statistics, the probability density function (PDF) of $\left| h_{SDi}\right|^2$ $(1 \le i \le M)$ is expressed as \cite{Men2015}
\begin{equation}
\begin{split}
f_{\left|h_{SDi}\right|^2}(x) &= \frac{{M!}}{{\left( {M - i} \right)!\left( {i - 1} \right)!}}\frac{1}{{{\lambda _{SD}}}}\sum\limits_{k = 0}^{i - 1} {\left( {\begin{array}{*{20}{c}}
{i - 1}\\
k
\end{array}} \right)} \\
& \hspace{0.8in} \times {\left( { - 1} \right)^k}{e^{ - x\left( {M - i + k+ 1} \right)/{\lambda _{SD}}}},
\end{split}
\label{pdfhsd}
\end{equation}
and its cumulative distribution function (CDF) can be written as
\begin{equation}
\begin{split}
{F_{{{\left| {{h_{SDi}}} \right|}^2}}}(x) &= \int\limits_0^x {{f_{{{\left| {{h_{SDi}}} \right|}^2}}}(t)dt} \\
&= \sum\limits_{k = 0}^{i - 1} {{\Phi _{k,i}}\left( {1 - {e^{ - x\left( {M - i + k + 1} \right)/{\lambda _{SD}}}}} \right)},
\end{split}
\label{cdfhsd}
\end{equation}
where ${\Phi _{k,i}} = \left( {\begin{array}{*{20}{c}}
{\begin{array}{*{20}{c}}
{i - 1}\\
k
\end{array}}
\end{array}} \right)\frac{{{{\left( 1 \right)}^k}M!}}{{\left( {M - i} \right)!\left( {i - 1} \right)!\left( {M - i + k + 1} \right)}}$ and ${\lambda _{SD}} = \mathbb{E}\left[ {{{\left| {{h_{SDi}}} \right|}^2}} \right]$, $\mathbb{E}\left[ \cdot \right]$ is the expectation operator.

\section{Performance analysis}
In this section, the performance analysis in terms of outage probability and system throughput is presented.

\subsection{Outage probability}
The outage probability of users $D_n$ and $D_m$ is analyzed through Theorem 1 and 2 as follows.

\textit{Theorem 1: Under Rayleigh fading channel, the outage probability of user $D_n$ can be expressed as}
\begin{equation}
P_{Out}^{(n)} = \sum\limits_{k = 0}^{n - 1} {{\Phi _{k,n}}} \left[ {1 - {e^{ - \beta \left( {M - n + k + 1} \right)/{\lambda _{SD}}}}} \right].
\label{poutn_fin}
\end{equation}
where ${\Phi _{k,n}} = \left( {\begin{array}{*{20}{c}}
{n - 1}\\
k
\end{array}} \right)\frac{{{{\left( { - 1} \right)}^k}M!}}{{\left( {M - n} \right)!\left( {n - 1} \right)!\left( {M - n + k + 1} \right)}}$, $\beta  = \max \left\{ {\alpha d_{SDn}^\theta ,\frac{{{\gamma _{thn}}d_{SDn}^\theta }}{{{a_n}{\gamma _0}}}} \right\}$, $\alpha = \frac{{{\gamma _{thm}} }}{{\left( {{a_m} - {a_n}{\gamma _{thm}}} \right){\gamma _0}}}$.

\begin{IEEEproof}
Since user $D_n$ needs to decode the signal of user $D_m$ first, the probability to characterize such an event can be formulated as
\begin{equation}
P_{Out}^{(n)} = 1 - \Pr \left( {{\gamma _{SDn \to m}} \ge {\gamma _{thm}}} \right)\Pr \left( {{\gamma _{SDn}} \ge {\gamma _{thn}}} \right).
\label{poutn_def}
\end{equation}

Substituting (\ref{gsdnm}) and (\ref{gsdn}) into (\ref{poutn_def}), the outage probability of user $D_n$ can be rewritten as
\begin{equation}
P_{Out}^{\left( n \right)} = 1 - \Pr \left( {{{\left| {{h_{SDn}}} \right|}^2} \ge \beta } \right) = \Pr \left( {{{\left| {{h_{SDn}}} \right|}^2} < \beta } \right).
\label{poutn_cal}
\end{equation}

It is important to note that the condition $\gamma_{thm} < \frac{a_m}{a_n}$ is used to obtain (\ref{poutn_cal}). The achieved result in (\ref{poutn_fin}) is attained by substituting (\ref{cdfhsd}) into (\ref{poutn_cal}).
\end{IEEEproof}

\textit{Theorem 2: The outage probability of user $D_m$ can be given by}
\begin{equation}
\begin{split}
P_{Out}^{\left( m \right)} &= \sum\limits_{k = 0}^{n - 1} {{\Phi _{k,n}}\left[ {1 - {e^{ - \alpha \left( {M - n + k + 1} \right)d_{SDn}^\theta /{\lambda _{SD}}}}} \right]} \\
& + \left\{ {1 - \sum\limits_{k = 0}^{n - 1} {{\Phi _{k,n}}\left[ {1 - {e^{ - \alpha \left( {M - n + k + 1} \right)d_{SDn}^\theta /{\lambda _{SD}}}}} \right]} } \right\} \\
& \times \sum\limits_{k = 0}^{m - 1} {{\Phi _{k,m}}\left[ {1 - {e^{ - \alpha \left( {M - m + k + 1} \right)d_{SDm}^\theta /{\lambda _{SD}}}}} \right]} \\
& \times \left[ {1 - {e^{ - \frac{{{\gamma _{thm}}}}{{{\gamma _0}}}\left( {\frac{{d_{RDm}^\theta }}{{{\lambda _{RDm}}}} + \frac{{d_{DnR}^\theta }}{{{\lambda _{DnR}}}}} \right)}}t{{\cal{K}}_1}(t)} \right],
\end{split}
\label{poutm_fin}
\end{equation}
where ${\Phi _{k,m}} = \left( {\begin{array}{*{20}{c}}
{m - 1}\\
k
\end{array}} \right)\frac{{{{\left( { - 1} \right)}^k}M!}}{{\left( {M - m} \right)!\left( {m - 1} \right)!\left( {M - m + k + 1} \right)}}$, $t = 2\sqrt {\frac{{d_{DnR}^\theta d_{RDm}^\theta {\gamma _{thm}}\left( {{\gamma _{thm}} + 1} \right)}}{{\gamma _0^2{\lambda _{DnR}}{\lambda _{RDm}}}}}$ and ${{\cal{K}}_1}(\cdot)$ denotes the $1^{st}$ - order modified Bessel function of the second kind.

\begin{IEEEproof}
Considering user $D_m$ with SC, an outage event occurs if and only if both the direct communication and the relaying communication are interrupted. Therefore, the outage probability of user $D_m$ can be shown as
\begin{equation}
\begin{split}
P_{Out}^{(m)} &= \Pr \left( {{\gamma _{SDn \to m}} < {\gamma _{thm}}} \right) + \left[ {1 - \Pr \left( {{\gamma _{SDn \to m}} < {\gamma _{thm}}} \right)} \right] \\
& \hspace{0.75in}\times \Pr \left( {{\gamma _{SDm}} < {\gamma _{thm}}} \right)\Pr \left( {{\gamma _{RDm}} < {\gamma _{thm}}} \right),
\end{split}
\label{poutm_def}
\end{equation}
where
\begin{equation}
\begin{split}
& \Pr \left( {{\gamma _{SDn \to m}} < {\gamma _{thm}}} \right) \\
&\hspace{0.6in} = \sum\limits_{k = 0}^{n - 1} {{\Phi _{k,n}}\left[ {1 - {e^{ - \alpha \left( {M - n + k + 1} \right)d_{SDn}^\theta /{\lambda _{SD}}}}} \right]},
\end{split}
\label{psdnm}
\end{equation}
\begin{equation}
\begin{split}
&\Pr \left( {{\gamma _{SDm}} < {\gamma _{thm}}} \right) \\
&\hspace{0.5in} = \sum\limits_{k = 0}^{n - 1} {{\Phi _{k,m}}\left[ {1 - {e^{ - \alpha \left( {M - m + k + 1} \right)d_{SDm}^\theta /{\lambda _{SD}}}}} \right]},
\end{split}
\label{psdm}
\end{equation}
and $\Pr \left( {{\gamma _{RDm}} < {\gamma _{thm}}} \right)$ can be calculated as shown in (\ref{prdm}) at the top of the next page. Note that the final result in (\ref{prdm}) is obtained after changing variable $u = {\gamma _0}x - d_{RDm}^\theta {\gamma _{thm}}$ and applying [\cite{Grad2007}, 3.324.1]. By substituting (\ref{psdnm}), (\ref{psdm}) and (\ref{prdm}) into (\ref{poutm_def}), the final expression of $P_{Out}^{(m)}$ is derived as (\ref{poutm_fin}).
\begin{small}
\begin{figure*}[!t]
\begin{equation}
\begin{split}
&\Pr \left( {{\gamma _{RDm}} < {\gamma _{thm}}} \right) \\
&= \Pr \left( {\frac{{\gamma _0^2{{\left| {{h_{RDm}}} \right|}^2}{{\left| {{h_{DnR}}} \right|}^2}}}{{{\gamma _0}d_{DnR}^\theta {{\left| {{h_{RDm}}} \right|}^2} + {\gamma _0}d_{RDm}^\theta {{\left| {{h_{DnR}}} \right|}^2} + d_{RDm}^\theta d_{DnR}^\theta }} < {\gamma _{thm}}} \right) \\
&= \Pr \left( {{{\left| {{h_{RDm}}} \right|}^2} < \frac{{d_{RDm}^\theta {\gamma _{thm}}}}{{{\gamma _0}}}} \right) + \Pr \left( {{{\left| {{h_{DnR}}} \right|}^2} < \frac{{{\gamma _{thm}}d_{DnR}^\theta \left( {{\gamma _0}{{\left| {{h_{RDm}}} \right|}^2} + d_{RDm}^\theta } \right)}}{{{\gamma _0}\left( {{\gamma _0}{{\left| {{h_{RDm}}} \right|}^2} - d_{RDm}^\theta {\gamma _{thm}}} \right)}},{{\left| {{h_{RDm}}} \right|}^2} \ge \frac{{d_{RDm}^\theta {\gamma _{thm}}}}{{{\gamma _0}}}} \right) \\
&= 1 - \int\limits_{\frac{{d_{RDm}^\theta {\gamma _{thm}}}}{{{\gamma _0}}}}^\infty  {\frac{1}{{{\lambda _{RDm}}}}{e^{ - \frac{x}{{{\lambda _{RDm}}}}}}{e^{ - \frac{{{\gamma _{thm}}d_{DnR}^\theta \left( {{\gamma _0}x + d_{RDm}^\theta } \right)}}{{{\gamma _0}\left( {{\gamma _0}x - d_{RDm}^\theta {\gamma _{thm}}} \right){\lambda _{DnR}}}}}}dx} \\
&  = 1 - {e^{ - \frac{{{\gamma _{thm}}}}{{{\gamma _0}}}\left( {\frac{{d_{RDm}^\theta }}{{{\lambda _{RDm}}}} + \frac{{d_{DnR}^\theta }}{{{\lambda _{DnR}}}}} \right)}}{2\sqrt {\frac{{d_{DnR}^\theta d_{RDm}^\theta {\gamma _{thm}}\left( {{\gamma _{thm}} + 1} \right)}}{{\gamma _0^2{\lambda _{DnR}}{\lambda _{RDm}}}}}}{K_1}\left(2\sqrt {\frac{{d_{DnR}^\theta d_{RDm}^\theta {\gamma _{thm}}\left( {{\gamma _{thm}} + 1} \right)}}{{\gamma _0^2{\lambda _{DnR}}{\lambda _{RDm}}}}}  \right),
\end{split}
\label{prdm}
\end{equation}

\hrulefill
\end{figure*}
\end{small}
\end{IEEEproof}

\subsection{System throughput}
In this subsection, we consider the throughput ($\tau$) in delay-sensitive transmission mode. Given the analytical results regarding the outage probability of users $D_n$ and $D_m$, the system throughput can be expressed relying on \cite{Liu2016} as below
\begin{equation}
\tau = \left(1-P_{Out}^{(n)}\right)R_n + \left(1-P_{Out}^{(m)}\right)R_m,
\end{equation}
where $P_{Out}^{(n)}$ and $P_{Out}^{(m)}$ are obtained from (\ref{poutn_fin}) and (\ref{poutm_fin}), respectively.

\section{Numerical results}

In this section, numerical results are provided to analyze the performance of the proposed scenario. Given this concern, our simulation focuses on the outage probability (OP) and the system throughput ($\tau$) metrics in the delay-sensitive transmission mode. 

Specifically, in the considered system, it is assumed that there exist six user nodes $(M = 6)$. In addition, the power allocation coefficients are set to be $a_m = 0.7$ and $a_n = 0.3$. Also, the data rates at nodes $D_m$ and $D_n$ are defined as $R_m = R_n = 1$ (bit/s/Hz). 
In particular, considering the relative distances between the nodes ($S$, $D_m$, $D_n$ and $R$), the values of ${d_{RDm}}$ and ${d_{DnDm}}$ can be calculated by a simple way as follows:
${d_{RDm}} = \sqrt {d_{DnDm}^2 + d_{DnR}^2 - 2{d_{DnDm}}{d_{DnR}}\cos {\alpha _1}}$, where $\alpha_1 = 40^o$ denotes the angle $\angle {D_m}{D_n}R$ 
, and ${d_{DnDm}} = \sqrt {d_{SDm}^2 + d_{SDn}^2 - 2{d_{SDm}}{d_{SDn}}\cos {\alpha _2}}$ with $\alpha_2 = 60^o$ represents the angle $\angle {D_m}S{D_n}$. 
To this end, we set the path loss exponent to be $\theta = 2$.

\begin{figure}[!t]
\centering
\includegraphics[scale = 0.38]{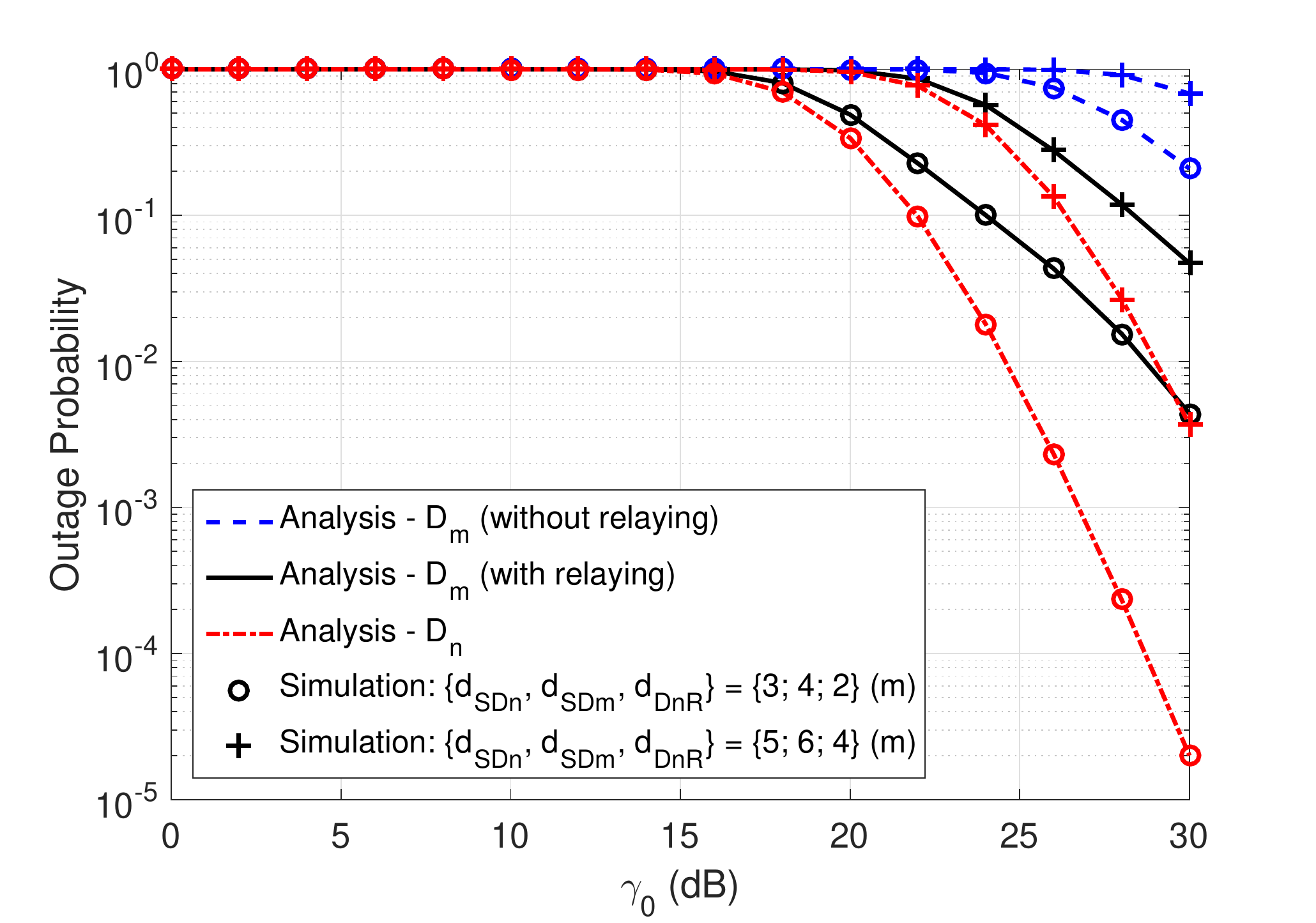}
\caption{Outage probability v.s. $\gamma_0$ with different values of the distances, where $m = 3$, $n = 6$.}
\label{fig:Poutd}
\end{figure}

\begin{figure}[!t]
\centering
\includegraphics[scale = 0.38]{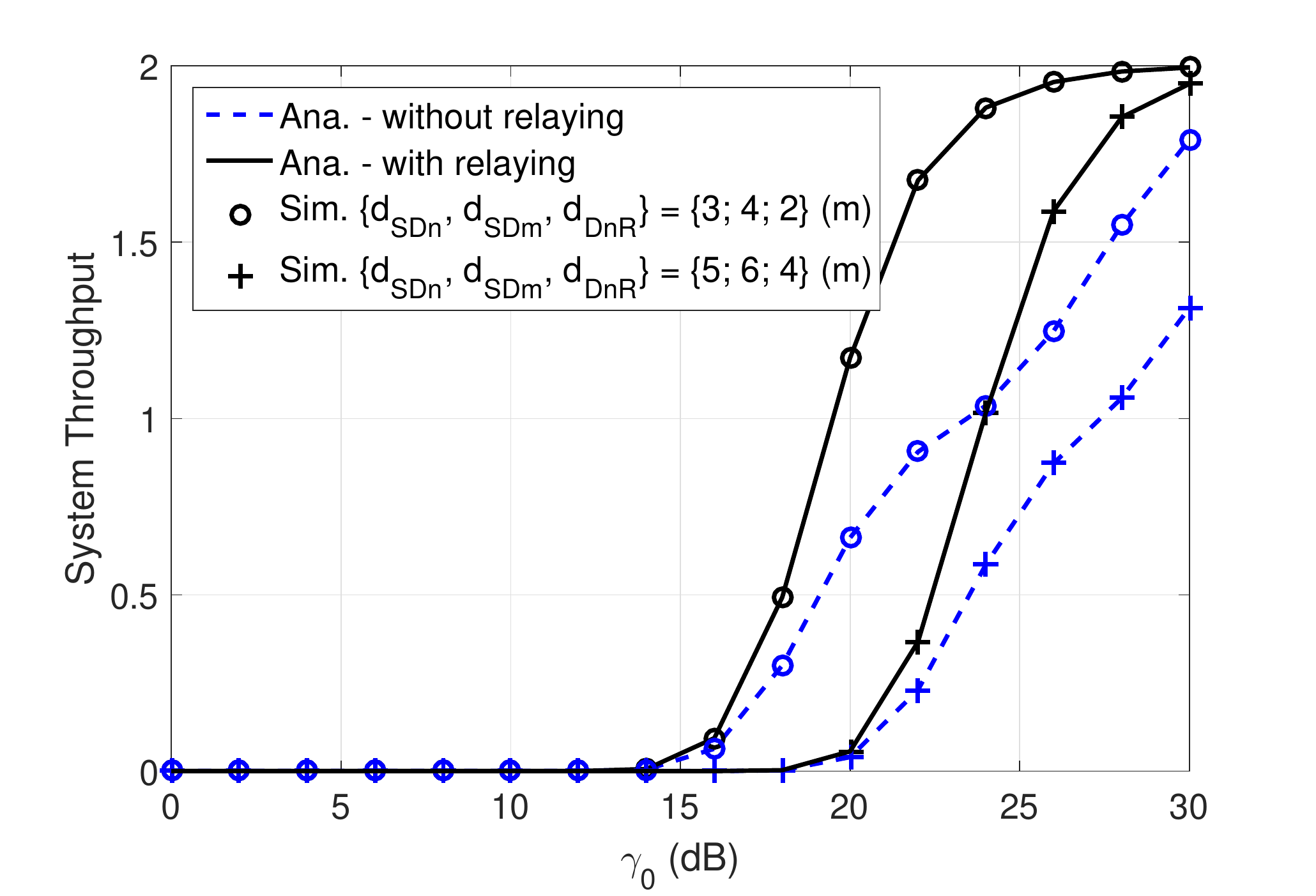}
\caption{System throughput v.s. $\gamma_0$ with different values of the distances, where $m = 3$, $n = 6$.}
\label{fig:taud}
\end{figure}

In Fig. \ref{fig:Poutd} and \ref{fig:taud}, the variation of the OP and the throughput with respect to average transmit SNR $\gamma_0$ is investigated in cases of different distance values, i.e. $d_{SDn}$, $d_{SDm}$, $d_{DnR}$, $d_{RDm}$, and $d_{DnDm}$. 
As observed in Fig. \ref{fig:Poutd}, it is visible that the increase in communication ranges results in the scale up of the OP. This implies a significant performance loss caused by the path loss. Also, the same phenomenon can be evaluated from Fig. \ref{fig:taud} in which a lower throughput is observed in the case that the longer distances are applied.
Furthermore, considering node $D_m$, one can observe from these two figures that the help of relay $R$ plays an important role in desirably improving the performance of the OP and the throughput.

\begin{figure}[!t]
\centering
\includegraphics[scale = 0.39]{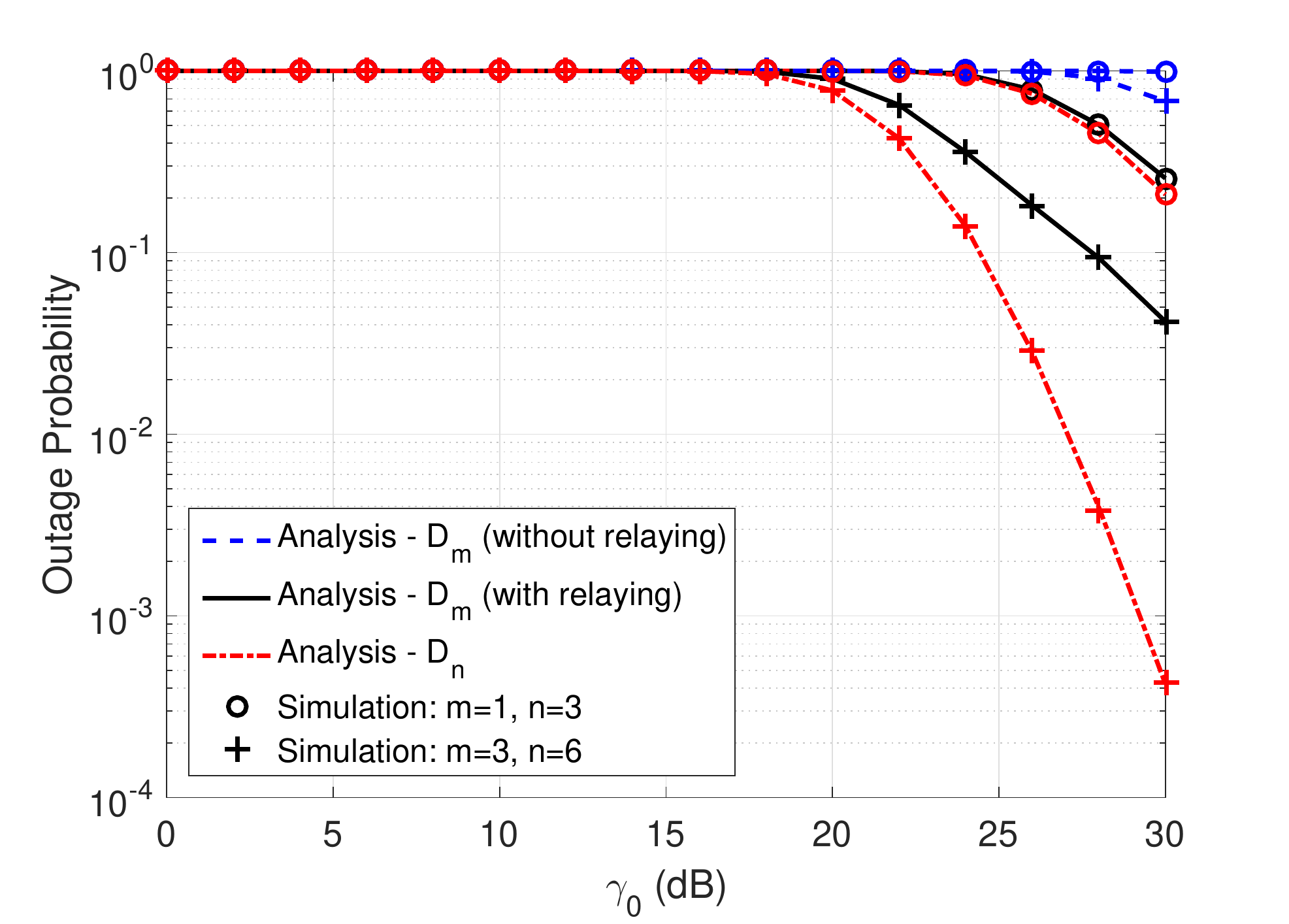}
\caption{Outage probability v.s. $\gamma_0$ with different values of $m$ and $n$, where $d_{SDn} = 4$(m), $d_{SDm} = 6$(m) and $d_{DnR} = 4$(m).}
\label{fig:Poutmn}
\end{figure}

\begin{figure}[!t]
\centering
\includegraphics[scale = 0.39]{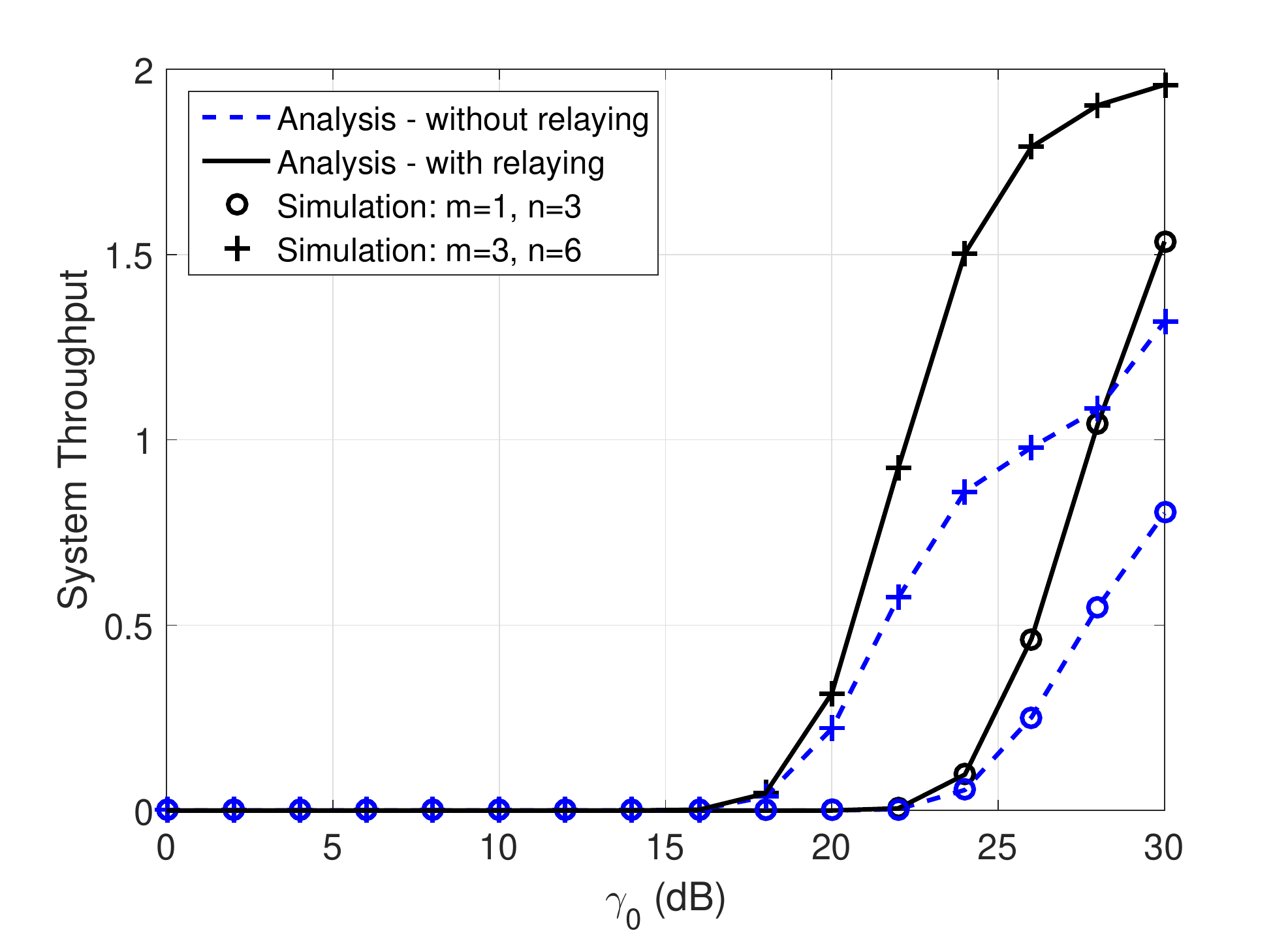}
\caption{System throughput v.s. $\gamma_0$ with different values of $m$ and $n$, where $d_{SDn} = 4$(m), $d_{SDm} = 6$(m) and $d_{DnR} = 4$(m).}
\label{fig:taumn}
\end{figure}

According to the principle of NOMA, differently selecting a pair of user nodes, i.e. $\{D_m, D_n\}$, leads to various changes in system performance.
To address this issue, Fig. \ref{fig:Poutmn} and Fig. \ref{fig:taumn} are plotted to identify how the selection affects the performance of the OP and the throughput, respectively. Given this concern, it is observed that the OP scales down whereas the througput scales up in the case that $m$ and $n$ are assigned with higher values. On this basis, it is suggested that $m$ and $n$ should be selected as large as possible to obtain the better performances.
Particularly, the provided discussion is confirmed by the fact that the analytical results are in a good agreement with the simulation results, as observed from all four figures.

\section{Conclusion}
In this work, we have raised the problem of limited user-to-user communications in NOMA systems, and then have proposed a new cooperative NOMA protocol to overcome such a problem.
 To evaluate the performance of the protocol, the closed-form expressions of the outage probability and the system throughput in delay-sensitive transmission mode have been derived. 
On this basis, the impact of some system parameters on the system performance has been investigated. 
Specifically, the analysis confirmed by simulation results show that the proposed scenario improves the system performance significantly. Finally, properly selecting a pair of users to perform NOMA is also suggested. 

\bibliographystyle{IEEEtran}
\bibliography{IEEEabrv,dung}

\end{document}